\documentclass[%
 amsmath,amssymb,
groupedaddress,%
twocolumn
]{revtex4}

\usepackage{graphicx}
\usepackage{dcolumn}
\usepackage{bm}

\usepackage[T1]{fontenc} 
\usepackage[applemac]{inputenc}
\usepackage{epstopdf}
\def\piezo{LiNbO$_\mathrm{3}$}

\graphicspath{{Fig_APL/}}
\usepackage{upgreek}

\usepackage{color}

\begin{document}

\preprint{AIP/123-QED}

\title{Fast acoustic tweezers for the two-dimensional manipulation of individual particles in microfluidic channels}

\author{S.B.Q. Tran}

\author{P. Marmottant}

\homepage{http://www-liphy.ujf-grenoble.fr/equipe/dyfcom/marmottant/Philippe_Marmottant}

\author{P. Thibault}

\email{Pierre.Thibault@ujf-grenoble.fr}

\affiliation{CNRS / Universit\'e Grenoble 1,  LIPhy UMR 5588, Grenoble, F-38401, France}

\date{\today}

\begin{abstract}
This paper presents a microfluidic device that  implements standing surface acoustic waves in order to handle single cells, droplets  and generally particles. The particles are moved in a very controlled manner by the two-dimensional drifting  of a standing wave array, using a slight frequency modulation of two ultrasound emitters around their resonance. These acoustic tweezers  allow any type of motion at velocities up to few $\times$10~mm/s while the device transparency is adapted for optical studies. The possibility of automation provides a critical step in the development of lab-on-a-chip cell sorters and it should find applications in biology, chemistry and engineering domains.

\end{abstract}

\maketitle
The recent progresses in lab on chip technologies have opened  numerous possibilities for medical diagnosis and biological tests. One issue is the adaptation of current technologies to microscale  for higher integration, cost reduction and parallel processing capabilities, as illustrated by the efforts to develop micro fluorescent activated cell sorters ($\upmu$FACS)  \cite{Fu1999,Franke2010}. Another issue is the manipulation of individual cells for \textit{in situ} diagnostics and optical characterization.  Surface acoustic wave (SAW) techniques are particularly promising for particle handling as they allow, when combined with a polydimethylsiloxane (PDMS)  microfluidic channel, optical access, local actuation and fast response time capabilities. 

In this Letter, we describe a device that can move particles down to the size of cells  \textit{and} precisely control their positions in a two-dimensional (2D) acoustic  cage. We show that the characteristics of these \textit{acoustic tweezers} are to be fast ($\times$10~mm/s) and to exert forces that are strong enough to  counteract friction from wall, or drag from hydrodynamic flows of reasonable velocities.  We present measurements of the particle trajectories obtained for an oscillatory driving motion, so as to understand the dynamics of slippage from the acoustic trap above a threshold velocity. Last, we propose an insight into the forces acting on particles in a microchannel under the action of SAW emission, providing evidence of their  3D action.

\begin{figure}[h]
	\begin{center}
	\includegraphics[width=9cm]{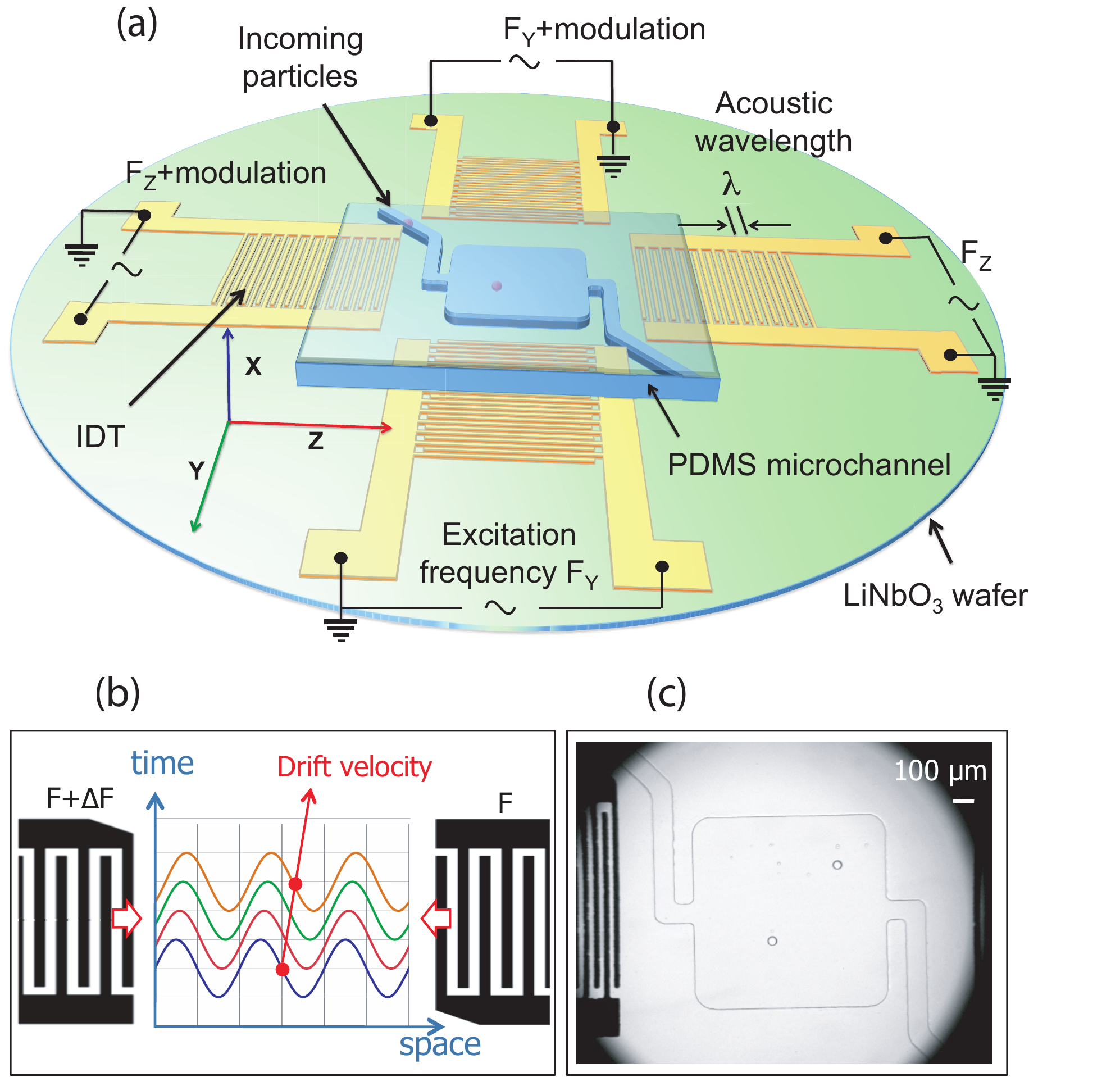} 
	\caption{\small a) Scheme of the acoustic device geometry. The flow direction is indicated by the arrow. A first pair of interdigitated transducers emits along the crystal strong axis Z which coincides with the main flow direction. A second pair emits along the perpendicular direction (Y~axis).  b) Principle of frequency modulation to move particles. c) Image of two oil droplets trapped in the acoustic chamber 1~mm wide and  50~$\upmu$m high.}
		\label{fig:device}
	\end{center}
\end{figure}

Friend \textit{et al.} \cite{Friend2011} have recently drawn a full panorama  of the potential offered by acoustic microfluidics based on SAW emission  in the case of standing waves (SSAW), propagating waves (PSAW) or both. Using PSAW, Franke \textit{et al.} have demonstrated the deviation of a continuous flow containing cells at high frequency  \cite{Franke2010}, and of droplets in a water in oil emulsion \cite{Franke2009}. Note that when using these pushing forces  the final position is not directly controlled because it depends on the surrounding fluid velocity. Using SSAW, Shi  \textit{et al.} were able to force bovine red blood cells particles to form lines   \cite{Shi2008} and obtained point-wise concentrations  \cite{Shi2009} of latex beads and \textit{E. Coli} cells with two interdigitated transducers (IDTs) in an orthogonal arrangement. In this case, it is believed that the observed patterns result from acoustic forces  due to SAW  leakage in the liquid. These forces pushing the particle  towards the pressure nodes,  this leads to a situation analog to bulk acoustic wave experiments  \cite{Manneberg2009,Wiklund2006}. Wood \textit{et al.} have also shown line wise \cite{Wood2008}  and point wise  \cite{Wood2009}  alignments of latex particles using opposite pairs of IDTs, the visualization of particles being obtained through a glass superstrate. In contrast with these experiments where the particles cannot be moved at will on predetermined locations,  Meng \textit{et al.}  \cite{Meng2011} have 
described a device  where micro-bubbles and breast cancers cells can be transported at will across the width of a microfluidic channel. 
The major improvement is the use of some phase shift between the two IDTs located on both part of to microfluidic channel to move the standing wave pattern, and hence the particles. Recently Ding \textit{et al.}  \cite{Ding2012} have proposed to implement chirped electrodes with a broadband response (in between 18.5 MHz and 37 MHz) to change the wavelength of a 2D standing wave pattern in both directions, and thus entrain particles.

In the present work, we have designed a device that cumulates the advantages of a 2D standing wave pattern to trap particles in some nodal positions and the continuous control of a small phase shift  to displace them, staying at the resonance frequency of the device. The experimental setup is schematized  on figure~\ref{fig:device}. The piezoelectric is a 2'', 500~$\upmu$m thick  \piezo~single crystal wafer, with X-cut section and Z as the main propagation direction ($K^2$ = 5.9\% in Z~direction, and 3.1\% in Y~direction \cite{Renaudin2009}). Electrodes forming the IDTs are produced by successive metal depositions (Ti/Au, 15~nm/100~nm) on the  \piezo~wafer with a standard lift-off process. They comprise 40~fingers electrode pairs with a constant pitch $\lambda$ of 100~$\upmu$m. The microfluidic channel, 50~$\upmu$m deep, is fabricated  in PDMS with soft-lithography using mold-replica techniques. It contains a flow focusing geometry \cite{Dollet2008} used to produce droplets, a main channel and a side branch~100~$\upmu$m wide leading to the main acoustic chamber 1~mm$\times$1~mm.  As the flow exerts an extra force on the particles to be counterbalanced by the IDT emission, it is aligned along~Z.  The channel is assembled  under microscope on the piezoelectric wafer, after it has been exposed for 10~s to an ozone plasma to promote adhesion between PDMS and \piezo\cite{Haubert2006}. The whole setup is heated to 65°C for two hours and clamped between plexiglas plates and two aluminum disks to avoid  the premature detachment of the channel. Visualization is made using a standard microscope (Macroscope Leica Z6) and particle trajectories are recorded using a  fast camera (Phantom Miro 4, Vision Research) and treated using ImageJ software (NIH, Bethesda, MD).

Two categories of particles are presented to illustrate the efficiency of the tweezers: oil in water droplets, and human red and white blood cells
 (hRBCs and hWBCs). The oil droplets are generated \textit{in situ} by injecting 50~cSt silicone oil (Sigma Aldrich) and de-ionized water with 10\% dishwashing surfactant (Dreft, Procter and Gamble) in the flow focus \cite{Garstecki2004}. The cells are obtained by successive concentrations using centrifugation and dilution in a buffer solution (PBS) for plasma removing. The IDTs are excited by a set of arbitrary waveform generators each connected to 40 or 50~dB amplifiers (Amplifier Research, model 75A250A and ENI, Models 320L, 325LA, 411LA). A calibration of the performance of each IDT as a function of frequency is initially performed by observing the deviation of some particles. The maxima are found at frequencies~$F$ in the range 34.5$-$37~MHz as expected  from the ratio between the velocity of the Rayleigh wave of the water loaded \piezo~to the IDT pitch \cite{Campbell1970}. A 1D or 2D standing wave pattern is created by exciting the IDT pairs at the same frequency and amplitude, for which the two counter-propagating waves compensate.  
 
Particle motion is achieved by the low frequency modulation of one IDT respective to its opposite. Periodic trajectories were obtained  by using modulations of the form  $F^{mod}=F+\Delta F \,g(2 \pi f_m t)$ where $\Delta F$ represents some frequency deviation and $f_m$ the modulation frequency in the periodic function $g$. Note that $\Delta F$ (of the order of Hz) is much smaller than~$F$, so that the device stays at resonance. This type of excitation automatically creates a dynamic phase shift $\varphi(t)$ with respect to the other electrode   such that  $\dot{\varphi}=2\pi \Delta F \,g(2 \pi f_m t)$ and thus a drift of the interference pattern at the velocity $v_d=\dot{\varphi} /2\pi \times \lambda / 2$. This contrasts with other works \cite{Meng2011}  where the motion was obtained by the progressive adjustment of the static relative phase between the two IDTs, or by an important modification of the frequency of both electrodes to change the wavelength \cite{Ding2012}.

 \begin{figure}[h]
	\begin{center}
	\includegraphics[width=8cm]{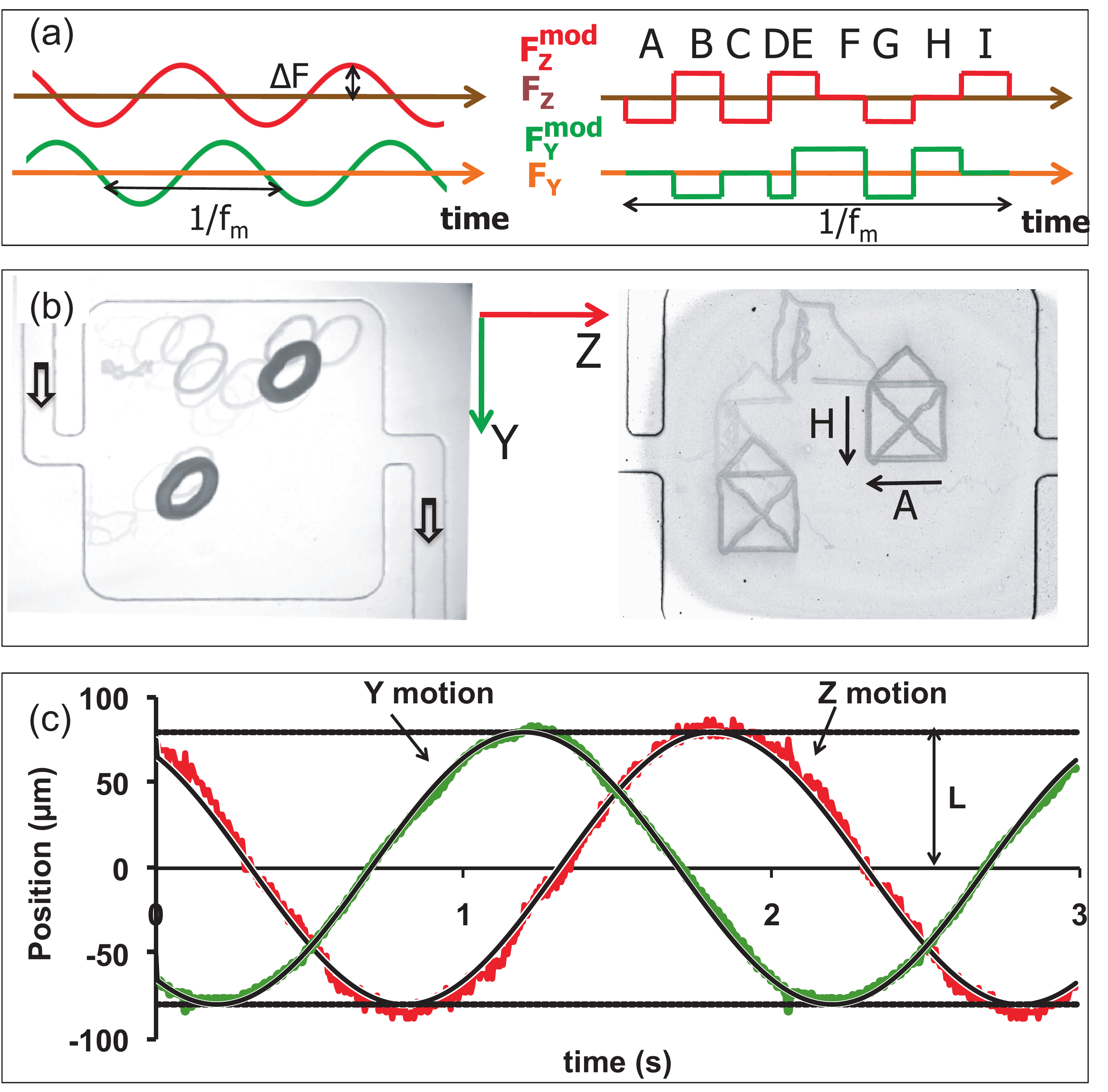} 
	\caption{\small a) Frequency versus time illustration of the modulation function. \textit{Left}: `elliptical' motion, $\Delta F$=5~Hz, $f_m$=0.5~Hz. \textit{Right} `house', $\Delta F$=32~Hz, $f_m$=1~Hz. The trajectories corresponding to modulations sequences A and H are shown by the arrows. b) Superimposed images of particles motions. IDT pairs frequency for the standing waves:  $F_Z$=36.8~MHz, $F_Y$=34.5~MHz. \textit{Left}: Silicone oil droplets. \textit{Right}: White blood cells.  c) Time evolution of the position of a silicone oil droplet position along flow (Z-axis) and transverse to the flow (Y-axis) and comparison with the model with fitting parameters $L=79.6~\upmu$m and $f_m$=0.5~Hz.}
		\label{fig:trajectories}
	\end{center}
\end{figure}

Figure~\ref{fig:trajectories} illustrates the results obtained with oil droplets that move under the action of two sine function modulations along the two directions~Y and~Z  (see also \textit{Supplementary Material}\cite{Note2011}).
The various droplets, differing by their diameters (40~$\upmu$m and 10~$\upmu$m for the smallest) follow identical elliptical  trajectories corresponding to Lissajous patterns. As a function of time,  a very good fit is obtained from the following parameter free model: $z=z_0+L \cos(2 \pi f_m t)$ and  $y=y_0+L \sin(2 \pi f_m t+\theta)$ with $L=\lambda \Delta F/ 4 \pi f_m$ half the amplitude of the course and $\theta$ some initial phase at the starting time of the SAW actuation. The figure also shows the trajectories of WBCs (diam. $\sim14~\upmu$m) with a modulation function programmed to describe a `house' shape. The control is still good, though some discrepancies can be seen. For RBCs, which sizes are smaller (6$-$8~$\upmu$m), the control gets more difficult at higher velocities and may depend on the remaining flow direction.

\begin{figure}[h]
	\begin{center}
	\includegraphics[width=8cm]{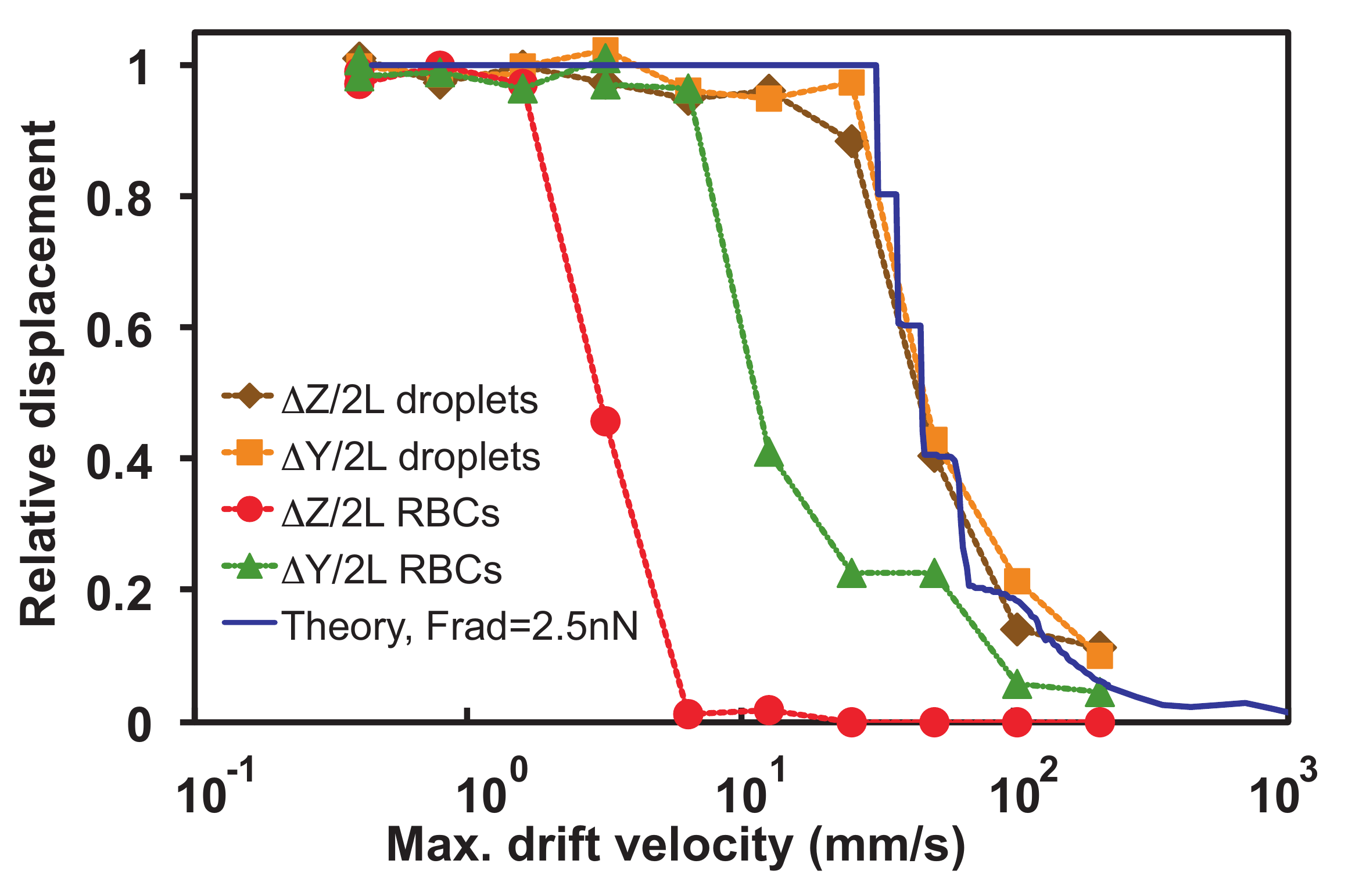} 
	\caption{\small Total amplitudes  of the periodic trajectories travelled by hRBCs and  $10~\upmu$m silicone oil droplets along the flow ($\Delta Z$) and transverse to the flow ($\Delta Y$) direction  as a function of the maximum calculated  drift velocity ($2\pi f_{m}L$). The values are normalized from  the amplitude of the total standing wave displacement $2L$. The continuous line showing plateaus corresponds to the numerical integration of Eq.~(\ref{eq:Josephson}) for oil droplets.}
		\label{fig:curves}
	\end{center}
\end{figure}

In order to determine the limit velocities at which particles can be moved, the distances traveled by the particles along (and transverse to) the main flow have been plotted in figure~\ref{fig:curves} as a function of the typical drift velocity $v_d$ for the sinusoidal modulation. Up to 25~mm/s, oil droplets have positions exactly predictable showing that  the  two degrees of freedom are independent. At larger velocities, the drifting amplitude is damped, since the viscous forces become comparable to the acoustic ones.  For RBCs, this maximum obtained velocity  reaches 7~mm/s in the flow transverse direction and is reduced to 1.5~mm/s along the main flow. 

The motion of the object can be predicted from the balance of forces at low Reynolds number:
\begin{equation}
F_{rad}\sin[4\pi (y-y_{0}(t))/\lambda]+6\pi \eta R \dot{y}=0	
\label{eq:Josephson}
\end{equation}
 where the first term is the radiation force on a object located at the position $y(t)$ in a standing wave (in the~Y or~Z direction) whose nodes move according to the position $y_{0}(t)=L\sin(2\pi f_{m}t)$, and where the second term is the Stokes drag at velocity~$\dot{y}$ in a fluid of viscosity~$\eta$.
The time integration of $\dot{y}$ gives an amplitude of displacement very close to that of figure~\ref{fig:curves}, with a sharp drop in displacement efficiency when the drifting velocity exceeds $v_{max}=F_{rad}/6\pi \eta R$, provided the radiation force is fitted by $F_{rad}=2.5$ nN for the $10~\upmu$m droplets (and $F_{rad}=0.1$ nN and 0.6 nN along the~Z and~Y direction for the RBCs, assuming a diameter of 8~$\upmu$m). 
We can  predict the acoustic pressure $P_{a}$ associated with theses forces to be $7.1\times 10^5$ Pa ($4.5\times 10^5$ and $9.8\times 10^5$ Pa for RBCs in~Z and~Y directions). These values for the pressures are obtained  using the expression by Yosioka \cite{Yosioka1955} for the radiation force as a function of acoustic pressure: $F_{rad}=\pi P_{a}^2V\Phi/2\lambda\rho c^2$, with $V=4\pi R^3/3$ and $\Phi$ a parameter that depends on the acoustic properties of objects and water
\cite{Note2012}.
  These predictions are of the order of magnitude expected for nanometric vibrations amplitudes  of the substrate \cite{Shilton2008} (as can be computed using $P_{a}=\rho c A \omega$, with $\rho$ the density of water, $c$ the sound velocity,  $A$ the vibration amplitude and $\omega$ the pulsation).

In an attempt to identify the parameters that influence the efficiency of the tweezers, we multiplied the experiments by varying particle sizes, particle composition, flow velocities, SAW amplitude and direction of emission. We observed that the proximity of the lateral channel walls (seen on Fig. 2(b)) could interfere with the particles motion due to undesired standing waves,  that originate from the reflexion of the incoming waves and that are fixed respective to the moving standing wave pattern.  
It was also found that the acoustic threshold necessary to initiate motion depended both on the particle size and its composition.  For a given size, the threshold is less for bubbles than for droplets, itself less than for cells. When the composition is fixed, the manipulation of smaller particles requires larger acoustic amplitudes. These observations are compatible with a mechanism where the forces at the origin of the acoustic tweezers are given by the gradient of acoustic pressure \cite{Yosioka1955}. 
However, some excessive acoustic amplitudes can also result in a   \textit{decrease} of the particle mobility as we now discuss.

\begin{figure}[h]
	\begin{center}
	\includegraphics[width=8cm]{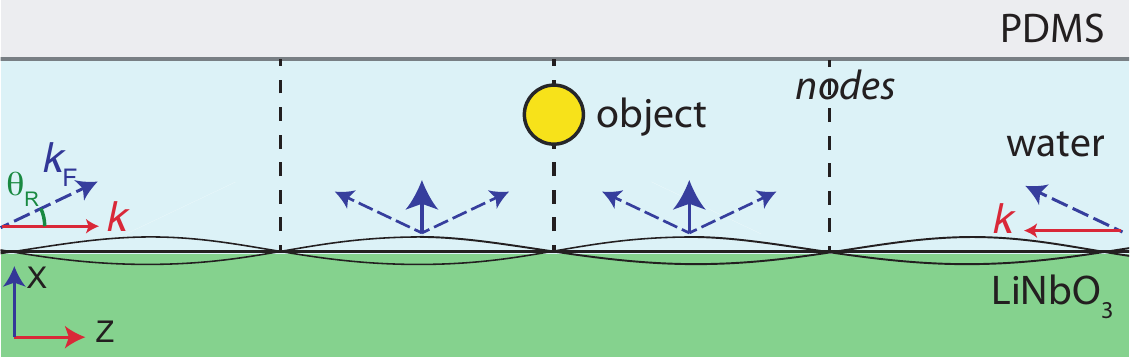} 
	\caption{\small Sketch of the 3D motion of a particle actuated by SAW propagation. The acoustic wave inside the liquid is at the same time a standing wave in a plane parallel to the surface (YZ~plane), and propagating along the perpendicular direction (X), since it is a superposition of two leaky waves, i.e. $\sin(k  z+k_{F}\sin\theta_{R}\;x-\omega t)+\sin(-k z+k_{F}\sin\theta_{R}\;x-\omega t)=2\cos(k z)\sin(k_{F}\sin\theta_{R}\;x-\omega t)$, where $k$ is the SAW wavevector, $\theta_{R}$ the Rayleigh angle,  and   $k_{F}=k/\cos\theta_{R}$ the acoustic wavevector in the liquid.  \label{fig:SSAW}}
	\end{center}
\end{figure}

As sketched on figure~\ref{fig:SSAW}, the creation of a standing wave from two counter propagating waves on the piezoelectric wafer also gives rise to a propagative component normal to the piezoelectric surface, which pushes the particles toward the PDMS walls. For objects whose dimensions exceed the channel height (large bubbles or droplets), this results in an increase of the friction forces at large acoustic  powers. Dewetting of the thin film in between the object and the PDMS top surface can even be observed for objects at rest, when the disjoining pressure is not large enough to counterbalance the acoustic forces. For objects smaller than the gap height  the mobility is also decreased, to an extent that depends on the top wall proximity and therefore on the amplitude of the radiation forces. This idea was verified by directly observing under a microscope  the accumulation of RBCs against the PDMS.  As a consequence, the search of an optimal acoustic power to manipulate particles with SAW involves the consideration of the 3D nature of the forces that are created.

In conclusion, we have developed and characterized acoustic tweezers that induce fast motion and precise positioning of single particles made of various materials.  We found that the action of the SSAW also involves normal radiative forces, and we have evidenced their effect on the particle mobility.
Eventually, we hope this work will contribute to the rapid development of efficient acoustic tweezers for cell manipulations, since it helps to adjust the amplitude of the acoustic force necessary to trap particles, depending on the desired drifting velocity.

The authors gratefully acknowledge financial support from ANR Microacoustics Grant ANR-08-JCJC-014-01.

\nocite{*} 

\begin{thebibliography}{0}
\expandafter\ifx\csname natexlab\endcsname\relax\def\natexlab#1{#1}\fi
\expandafter\ifx\csname bibnamefont\endcsname\relax
  \def\bibnamefont#1{#1}\fi
\expandafter\ifx\csname bibfnamefont\endcsname\relax
  \def\bibfnamefont#1{#1}\fi
\expandafter\ifx\csname citenamefont\endcsname\relax
  \def\citenamefont#1{#1}\fi
\expandafter\ifx\csname url\endcsname\relax
  \def\url#1{\texttt{#1}}\fi
\expandafter\ifx\csname urlprefix\endcsname\relax\def\urlprefix{URL }\fi
\providecommand{\bibinfo}[2]{#2}
\providecommand{\eprint}[2][]{\url{#2}}

\end{thebibliography}


\begin{mcitethebibliography}{23}
\providecommand*{\natexlab}[1]{#1}
\providecommand*{\mciteSetBstSublistMode}[1]{}
\providecommand*{\mciteSetBstMaxWidthForm}[2]{}
\providecommand*{\mciteBstWouldAddEndPuncttrue}
  {\def\EndOfBibitem{\unskip.}}
\providecommand*{\mciteBstWouldAddEndPunctfalse}
  {\let\EndOfBibitem\relax}
\providecommand*{\mciteSetBstMidEndSepPunct}[3]{}
\providecommand*{\mciteSetBstSublistLabelBeginEnd}[3]{}
\providecommand*{\EndOfBibitem}{}
\mciteSetBstSublistMode{f}
\mciteSetBstMaxWidthForm{subitem}
{(\emph{\alph{mcitesubitemcount}})}
\mciteSetBstSublistLabelBeginEnd{\mcitemaxwidthsubitemform\space}
{\relax}{\relax}

\bibitem[Fu \emph{et~al.}(1999)Fu, Spence, Scherer, Arnold, and Quake]{Fu1999}
A.~Y. Fu, C.~Spence, A.~Scherer, F.~H. Arnold and S.~R. Quake, \emph{Nat
  Biotech}, 1999, \textbf{17}, 1109--1111\relax
\mciteBstWouldAddEndPuncttrue
\mciteSetBstMidEndSepPunct{\mcitedefaultmidpunct}
{\mcitedefaultendpunct}{\mcitedefaultseppunct}\relax
\EndOfBibitem
\bibitem[Franke \emph{et~al.}(2010)Franke, Braunmuller, Schmid, Wixforth, and
  Weitz]{Franke2010}
T.~Franke, S.~Braunmuller, L.~Schmid, A.~Wixforth and D.~A. Weitz, \emph{Lab
  Chip}, 2010, \textbf{10}, 789--794\relax
\mciteBstWouldAddEndPuncttrue
\mciteSetBstMidEndSepPunct{\mcitedefaultmidpunct}
{\mcitedefaultendpunct}{\mcitedefaultseppunct}\relax
\EndOfBibitem
\bibitem[Friend and Yeo(2011)]{Friend2011}
J.~Friend and L.~Y. Yeo, \emph{Rev. Mod. Phys.}, 2011, \textbf{83},
  647--704\relax
\mciteBstWouldAddEndPuncttrue
\mciteSetBstMidEndSepPunct{\mcitedefaultmidpunct}
{\mcitedefaultendpunct}{\mcitedefaultseppunct}\relax
\EndOfBibitem
\bibitem[Franke \emph{et~al.}(2009)Franke, Abate, Weitz, and
  Wixforth]{Franke2009}
T.~Franke, A.~R. Abate, D.~A. Weitz and A.~Wixforth, \emph{Lab Chip}, 2009,
  \textbf{9}, 2625\relax
\mciteBstWouldAddEndPuncttrue
\mciteSetBstMidEndSepPunct{\mcitedefaultmidpunct}
{\mcitedefaultendpunct}{\mcitedefaultseppunct}\relax
\EndOfBibitem
\bibitem[Shi \emph{et~al.}(2008)Shi, Mao, Ahmed, Colletti, and Huang]{Shi2008}
J.~Shi, X.~Mao, D.~Ahmed, A.~Colletti and T.~J. Huang, \emph{Lab on a Chip},
  2008, \textbf{8}, 221--223\relax
\mciteBstWouldAddEndPuncttrue
\mciteSetBstMidEndSepPunct{\mcitedefaultmidpunct}
{\mcitedefaultendpunct}{\mcitedefaultseppunct}\relax
\EndOfBibitem
\bibitem[Shi \emph{et~al.}(2009)Shi, Ahmed, Mao, Lin, Lawit, and
  Huang]{Shi2009}
J.~Shi, D.~Ahmed, X.~Mao, S.-C.~S. Lin, A.~Lawit and T.~J. Huang, \emph{Lab
  Chip}, 2009, \textbf{9}, 2890--2895\relax
\mciteBstWouldAddEndPuncttrue
\mciteSetBstMidEndSepPunct{\mcitedefaultmidpunct}
{\mcitedefaultendpunct}{\mcitedefaultseppunct}\relax
\EndOfBibitem
\bibitem[Manneberg \emph{et~al.}(2009)Manneberg, Vanherberghen, Onfelt, and
  Wiklund]{Manneberg2009}
O.~Manneberg, B.~Vanherberghen, B.~Onfelt and M.~Wiklund, \emph{Lab Chip},
  2009, \textbf{9}, 833--837\relax
\mciteBstWouldAddEndPuncttrue
\mciteSetBstMidEndSepPunct{\mcitedefaultmidpunct}
{\mcitedefaultendpunct}{\mcitedefaultseppunct}\relax
\EndOfBibitem
\bibitem[Wiklund and Hertz(2006)]{Wiklund2006}
M.~Wiklund and H.~M. Hertz, \emph{Lab Chip}, 2006, \textbf{6}, 1279--1292\relax
\mciteBstWouldAddEndPuncttrue
\mciteSetBstMidEndSepPunct{\mcitedefaultmidpunct}
{\mcitedefaultendpunct}{\mcitedefaultseppunct}\relax
\EndOfBibitem
\bibitem[{Wood} \emph{et~al.}(2008){Wood}, {Evans}, {Cunningham}, {O'Rorke},
  {W{\"a}lti}, and {Davies}]{Wood2008}
C.~D. {Wood}, S.~D. {Evans}, J.~E. {Cunningham}, R.~{O'Rorke}, C.~{W{\"a}lti}
  and A.~G. {Davies}, \emph{Applied Physics Letters}, 2008, \textbf{92},
  044104\relax
\mciteBstWouldAddEndPuncttrue
\mciteSetBstMidEndSepPunct{\mcitedefaultmidpunct}
{\mcitedefaultendpunct}{\mcitedefaultseppunct}\relax
\EndOfBibitem
\bibitem[{Wood} \emph{et~al.}(2009){Wood}, {Cunningham}, {O'Rorke},
  {W{\"a}lti}, {Linfield}, {Davies}, and {Evans}]{Wood2009}
C.~D. {Wood}, J.~E. {Cunningham}, R.~{O'Rorke}, C.~{W{\"a}lti}, E.~H.
  {Linfield}, A.~G. {Davies} and S.~D. {Evans}, \emph{Applied Physics Letters},
  2009, \textbf{94}, 054101\relax
\mciteBstWouldAddEndPuncttrue
\mciteSetBstMidEndSepPunct{\mcitedefaultmidpunct}
{\mcitedefaultendpunct}{\mcitedefaultseppunct}\relax
\EndOfBibitem
\bibitem[Meng \emph{et~al.}(2011)Meng, Cai, Zhang, Niu, Jin, Yan, Wu, Wang, and
  Zheng]{Meng2011}
L.~Meng, F.~Cai, Z.~Zhang, L.~Niu, Q.~Jin, F.~Yan, J.~Wu, Z.~Wang and H.~Zheng,
  \emph{Biomicrofluidics}, 2011, \textbf{5}, 044104\relax
\mciteBstWouldAddEndPuncttrue
\mciteSetBstMidEndSepPunct{\mcitedefaultmidpunct}
{\mcitedefaultendpunct}{\mcitedefaultseppunct}\relax
\EndOfBibitem
\bibitem[Ding \emph{et~al.}(2012)Ding, Lin, Kiraly, Yue, Li, Chiang, Shi,
  Benkovic, and Huang]{Ding2012}
X.~Ding, S.-C.~S. Lin, B.~Kiraly, H.~Yue, S.~Li, I.-K. Chiang, J.~Shi, S.~J.
  Benkovic and T.~J. Huang, \emph{Proceedings of the National Academy of
  Sciences}, 2012, \textbf{109}, 11105--11109\relax
\mciteBstWouldAddEndPuncttrue
\mciteSetBstMidEndSepPunct{\mcitedefaultmidpunct}
{\mcitedefaultendpunct}{\mcitedefaultseppunct}\relax
\EndOfBibitem
\bibitem[Renaudin \emph{et~al.}(2009)Renaudin, Sozanski, Verbeke, Zhang,
  Tabourier, and Druon]{Renaudin2009}
A.~Renaudin, J.-P. Sozanski, B.~Verbeke, V.~Zhang, P.~Tabourier and C.~Druon,
  \emph{Sensors and Actuators B: Chemical}, 2009, \textbf{138}, 374--382\relax
\mciteBstWouldAddEndPuncttrue
\mciteSetBstMidEndSepPunct{\mcitedefaultmidpunct}
{\mcitedefaultendpunct}{\mcitedefaultseppunct}\relax
\EndOfBibitem
\bibitem[Dollet \emph{et~al.}(2008)Dollet, van Hoeve, Raven, Marmottant, and
  Versluis]{Dollet2008}
B.~Dollet, W.~van Hoeve, J.-P. Raven, P.~Marmottant and M.~Versluis,
  \emph{Phys. Rev. Lett.}, 2008, \textbf{100}, 034504--4\relax
\mciteBstWouldAddEndPuncttrue
\mciteSetBstMidEndSepPunct{\mcitedefaultmidpunct}
{\mcitedefaultendpunct}{\mcitedefaultseppunct}\relax
\EndOfBibitem
\bibitem[Haubert \emph{et~al.}(2006)Haubert, Drier, and Beebe]{Haubert2006}
K.~Haubert, T.~Drier and D.~Beebe, \emph{Lab Chip}, 2006, \textbf{6},
  1548--1549\relax
\mciteBstWouldAddEndPuncttrue
\mciteSetBstMidEndSepPunct{\mcitedefaultmidpunct}
{\mcitedefaultendpunct}{\mcitedefaultseppunct}\relax
\EndOfBibitem
\bibitem[Garstecki \emph{et~al.}(2004)Garstecki, Gitlin, DiLuzio, Whitesides,
  Kumacheva, and Stone]{Garstecki2004}
P.~Garstecki, I.~Gitlin, W.~DiLuzio, G.~M. Whitesides, E.~Kumacheva and H.~A.
  Stone, \emph{Appl. Phys. Lett.}, 2004, \textbf{85}, 2649--2651\relax
\mciteBstWouldAddEndPuncttrue
\mciteSetBstMidEndSepPunct{\mcitedefaultmidpunct}
{\mcitedefaultendpunct}{\mcitedefaultseppunct}\relax
\EndOfBibitem
\bibitem[{Campbell} and {Jones}(1970)]{Campbell1970}
J.~J. {Campbell} and W.~R. {Jones}, \emph{Journal of Applied Physics}, 1970,
  \textbf{41}, 2796--2801\relax
\mciteBstWouldAddEndPuncttrue
\mciteSetBstMidEndSepPunct{\mcitedefaultmidpunct}
{\mcitedefaultendpunct}{\mcitedefaultseppunct}\relax
\EndOfBibitem
\bibitem[Not()]{Note2011}
Supplementary material available at http://?? with 4 movies illustrating the
  acoustic manipulation of droplets and biological cells in the microfluidic
  chamber\relax
\mciteBstWouldAddEndPuncttrue
\mciteSetBstMidEndSepPunct{\mcitedefaultmidpunct}
{\mcitedefaultendpunct}{\mcitedefaultseppunct}\relax
\EndOfBibitem
\bibitem[Yosioka and Kawasima(1955)]{Yosioka1955}
K.~Yosioka and Y.~Kawasima, \emph{Acustica}, 1955, \textbf{5}, 167\relax
\mciteBstWouldAddEndPuncttrue
\mciteSetBstMidEndSepPunct{\mcitedefaultmidpunct}
{\mcitedefaultendpunct}{\mcitedefaultseppunct}\relax
\EndOfBibitem
\bibitem[Not()]{Note2012}
$\Phi =(5\rho _1-2\rho )/(2\rho _1+\rho )-\rho c^2/\rho _1 c_{1}^2$, where the
  density $\rho _{1}$ and speed of sound $c_{1}$ within the object are
  respectively 970~kg/m$^3$ and 997~m/s for silicone oil \cite {Bertin2011}, or
  1099~kg/m$^3$ and 1646~m/s for RBCs in isotonic solution \cite {Weiser1982},
  taking for water $\rho $~=~1000~kg/m$^3$ and $c$~=~1500~m/s\relax
\mciteBstWouldAddEndPuncttrue
\mciteSetBstMidEndSepPunct{\mcitedefaultmidpunct}
{\mcitedefaultendpunct}{\mcitedefaultseppunct}\relax
\EndOfBibitem
\bibitem[{Shilton} \emph{et~al.}(2008){Shilton}, {Tan}, {Yeo}, and
  {Friend}]{Shilton2008}
R.~{Shilton}, M.~K. {Tan}, L.~Y. {Yeo} and J.~R. {Friend}, \emph{Journal of
  Applied Physics}, 2008, \textbf{104}, 014910\relax
\mciteBstWouldAddEndPuncttrue
\mciteSetBstMidEndSepPunct{\mcitedefaultmidpunct}
{\mcitedefaultendpunct}{\mcitedefaultseppunct}\relax
\EndOfBibitem
\bibitem[Bertin(2011)]{Bertin2011}
N.~Bertin, \emph{Ph.D. thesis}, Universit\'e Bordeaux I, 2011\relax
\mciteBstWouldAddEndPuncttrue
\mciteSetBstMidEndSepPunct{\mcitedefaultmidpunct}
{\mcitedefaultendpunct}{\mcitedefaultseppunct}\relax
\EndOfBibitem
\bibitem[Weiser and Apfel(1982)]{Weiser1982}
M.~A.~H. Weiser and R.~E. Apfel, \emph{The Journal of the Acoustical Society of
  America}, 1982, \textbf{71}, 1261--1268\relax
\mciteBstWouldAddEndPuncttrue
\mciteSetBstMidEndSepPunct{\mcitedefaultmidpunct}
{\mcitedefaultendpunct}{\mcitedefaultseppunct}\relax
\EndOfBibitem
\end{mcitethebibliography}
\providecommand*{\mcitethebibliography}{\thebibliography}
\csname @ifundefined\endcsname{endmcitethebibliography}
{\let\endmcitethebibliography\endthebibliography}{}

\end{document}